\documentclass[aps,prl,preprint,groupedaddress,nofootinbib]{revtex4}

\usepackage{graphicx}

\begin{document}

\title{Collision dynamics of two $^{238}$U atomic nuclei}

\author{C\'edric Golabek}
\affiliation{
GANIL (IN2P3/CNRS - DSM/CEA), BP 55027, F-14076 Caen Cedex 5, France.
}

\author{C\'edric Simenel}
 \affiliation{
CEA, Centre de Saclay, IRFU/Service de Physique Nucl\'eaire, F-91191 Gif-sur-Yvette, France.}

\date{\today}

\begin{abstract}

Collisions of actinide nuclei form, during very short times of few $10^{-21}$~s, 
the heaviest ensembles of interacting nucleons available on Earth. 
Such very heavy ions collisions have been proposed as an alternative way to produce heavy and superheavy elements.
These collisions are also used to produce super-strong electric fields by the huge number of interacting protons to test spontaneous positron-electron ($e^+e^-$) pair emission predicted by the quantum electrodynamics theory. 
The time-dependent Hartree-Fock theory which is a fully microscopic quantum approach is used to study collision dynamics of two $^{238}$U atomic nuclei. In particular, the role of nuclear deformation on collision time and on reaction mechanisms such as nucleon transfer is emphasized.
These calculations are pessimistic in terms of transfermium elements ($Z>100$) production.
However, the highest collision times  ($\sim4\times10^{-21}$~s at 1200~MeV) should allow experimental signature of spontaneous $e^+e^-$ emission in case of bare uranium ions. Surprisingly, we also observe ternary fission due to purely dynamical effects. 
\end{abstract}


\maketitle

The study of nuclei with more than 100 protons is strongly motivated by the desire to understand quantum mechanics at the scale of few femtometers as there existence relies only on quantum shell effects. Indeed, in a purely classical world, i.e., without shell structure, transfermium nuclei would undergo fission within about 10$^{-20}$~s due to the strong Coulomb repulsion between their protons. 
In one hand, SHEs are searched to localize the next island of stability in the top of the nuclear chart~\cite{oga99,hof00,cwi05,her06,oga06,mor07,hof07,mor08}. 
In the other hand, SHEs provide a crucial test to modern atomic models as their chemical properties might deviate from their homologue elements in the periodic table due to strong relativistic effects on the valence electron shells~\cite{eic07}.
Recently, SHEs have been synthesized through ''cold'' fusion reactions based on closed shell target 
nuclei~\cite{hof00,mor07} and with ''hot'' fusion reactions involving actinide targets~\cite{oga99,oga06,hof07} where nuclei up to $Z=118$ have been produced~\cite{oga06}. However, the decay chains of nuclei formed by hot fusion do not populate presently known nuclei and a ''blank spot'' exists in the nuclear chart around $Z=105$ and $N=160$. Modern experimental technics might be used to explore this region with multinucleon transfer between actinides~\cite{zag06}, and, thus, deserve theoretical investigations which are addressed in this letter. 

Our study also deals with the possibility to produce $e^+e^-$ spontaneous emission in such a collision~\cite{rei81,gre83,ack08}. 
No experimental evidence of this process has been obtained so far~\cite{ahm99}. 
One limiting factor is the Pauli blocking effect due to an occupation of the final state by surrounding electrons.
However, future facilities like the GSI-FAIR project should be able to get rid of this limitation using bare uranium-uranium merged-beam collisions.
Reliable predictions of collision times are needed to optimize the energy of the reaction to get the longest sticking times between the fragments. 
Recent theoretical calculations based on the time-dependent Dirac equation~\cite{ack08} show that two bare $^{238}$U need to stick together during at least $2.10^{-21}$~s to allow observation of spontaneous positron emission. 
Although no pocket exists in the nucleus-nucleus potential of this system~\cite{ber90,sim09}, nuclear attraction reduces Coulomb repulsion and dissipation mechanisms such as evolution of nuclear shapes may delay the separation of the system~\cite{zag06}. 
Recently, delay times in this reaction was searched analyzing kinetic energy loss and mass transfer~\cite{gol08}. 
Theoretically, the complexity of reaction mechanisms and the high number of degrees of freedom to be included motivate the use of microscopic approaches. First dynamical microscopic calculations of $^{238}$U+$^{238}$U have been performed recently thanks to the Quantum Molecular Dynamics (QMD) model~\cite{tia08}. 
Though a major step forward has been done in terms of predictive power with these calculations, improvements are mandatory for a more realistic description of collision dynamics. In particular, the strong ground state deformation of $^{238}$U is not included and nucleon wave functions are constrained to be Gaussian wave packets. 
In addition, the Pauli principle is only approximately treated in QMD. 

In the present work, we overcome these limitations by authorizing all possible spatial forms of the nucleon wave functions. 
The time dependent Hartree-Fock (TDHF) theory proposed by Dirac~\cite{dir30} is used with a Skyrme energy density functional (EDF) modeling nuclear interactions between nucleons~\cite{sky56}. The EDF is the only phenomenological ingredient of the model, as it has been adjusted on nuclear structure properties like infinite nuclear matter and radii and masses of few doubly magic nuclei~\cite{cha98}. The main approximation of this theory is to constrain the many-body wave function to be an antisymetrized independent particles state at any time. It ensures an exact treatment of the Pauli principle during time evolution. Though TDHF does not include two-body collision term, it is expected to treat correctly one-body dissipation which is known to drive low energy reaction mechanisms as Pauli blocking prevents nucleon-nucleon collisions.
Inclusion of pairing correlations responsible for superfluidity in nuclei have been done only recently with a full Skyrme EDF within the time dependent Hartree-Fock-Bogolyubov theory to study pairing vibrations in nuclei~\cite{ave08}. However, realistic applications to heavy ions collisions are not yet achieved and are beyond the scope of this work.
At initial time, the nuclei are in their Hartree-Fock ground state~\cite{har28,foc30}
allowing for a fully consistent treatment of nuclear structure and dynamics.  

The TDHF equation can be written as a Liouville-Von Neumann equation 
\begin{equation}
i\hbar \frac{\partial}{\partial t} \rho = \left[h[\rho],\rho\right]
\label{eq:tdhf}
\end{equation}
where $\rho$ is the one body density matrix associated to the total independent particles state with elements 
\begin{equation}
\rho(\mathbf{r} sq, \mathbf{r'}s'q') = \sum_{i=1}^{A_1+A_2} \varphi_i(\mathbf{r} sq)\varphi_i^*(\mathbf{r'}s'q')
\end{equation}
where $A_1$ and $A_2$ are the number of nucleons in the nuclei.
The sum runs over all occupied single particle wave functions $\varphi_i$ and $\mathbf{r}$, $s$ and $q$ denote the nucleon position, spin and isospin respectively.
The Hartree-Fock single particle Hamiltonian $h[\rho]$ is related to the EDF, noted $E[\rho]$, which depends on time-even and time-odd local densities~\cite{eng75} by its first derivative
\begin{equation}
h[\rho](\mathbf{r} sq, \mathbf{r'}s'q') = \frac{\delta E[\rho]}{\delta \rho(\mathbf{r'} s'q', \mathbf{r} sq)}.
\end{equation}

First applications of TDHF to nuclear collisions were restricted to calculations in one dimension~\cite{bon76}.
Recent increase of computational power allowed realistic TDHF calculations of heavy ions collisions in 3 dimensions with modern Skyrme functionals including spin-orbit term~\cite{kim97,uma06a,mar06}. 
In this work,  Eq.~\ref{eq:tdhf} is solved iteratively in time on a spatial grid with a plane of symmetry (the collision plane) using the {\textsc{tdhf3d}} code built by P. Bonche and coworkers with the SLy4$d$ parameterization of the Skyrme EDF~\cite{kim97}.
The lattice spacing is $\Delta~x=0.8$~fm and the time step is $\Delta~t=1.5\times10^{-24}$~s  (see also Ref.~\cite{sim08b} for more practical details of the numerical implementation). 
This code has been extensively used to study heavy ions fusion~\cite{sim01,sim08,was08,iwa08,was09,sim09}. In particular, it reproduces average fusion barriers very well without any additional parameter than the EDF ones, i.e., with no input from reaction mechanisms~\cite{sim08,was08}. Recent calculations also indicate that TDHF can be used to study the fusion hindrance phenomenon observed in heavy quasi-symmetric systems~\cite{sim09}. The latter is encouraging to apply TDHF to collisions with heavier reactants.

\begin{figure}
\includegraphics[width=85mm]{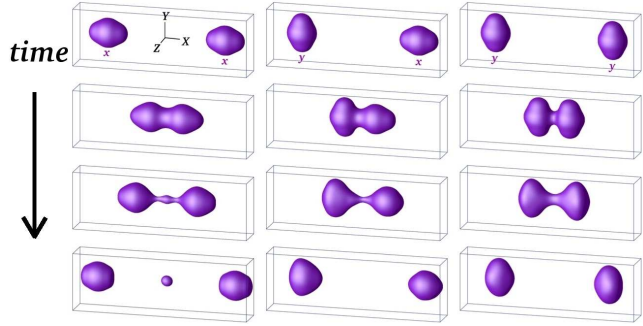}
\caption{Isodensities at half the saturation density in $^{238}$U+$^{238}$U central collision at a center of mass energy $E_{CM}=900$~MeV.
 Evolutions associated to the three initial configurations $xx$, $yx$ and $yy$ are plotted in the left, middle and right column respectively. Snapshots are given at times $t=0$, 15, 27 and $42\times10^{-22}$~s from top to bottom.}
\label{fig:densities}
\end{figure}

The $^{238}$U nucleus exhibits a prolate deformation with a symmetry axis in its ground state. 
The effect of this deformation on collision is investigated in four configurations ($xx$, $yx$, $yy$ and $yz$) associated to different initial orientations. 
The letters $x$, $y$ and $z$ denote the symmetry axis of the nuclei which collide along the $x$ axis (see, e.g., top of Figure~\ref{fig:densities}). We focus on central collisions as they lead to the most dissipative reactions with the longest collision times. 

First, we analyze the fragments produced in exit channels. Strictly speaking, they are {\it primary} fragments as they might decay by statistical fission. This decay is not studied here as it occurs on a much longer time scale than the collision itself.  
The importance of initial orientation on reaction mechanism is clearly seen in Figure~\ref{fig:densities}. Snapshots of isodensities at half saturation density, i.e., $\rho_0/2=0.08$~fm$^{-3}$, are plotted for the $xx$, $yx$ and $yy$ configurations at a center of mass energy $E_{CM}=900$~MeV.
The $yy$~configuration gives two symmetric fragments because, in this particular collision, the $x=0$ plane is a plane of symmetry, although nucleon transfer is still possible in such a symmetric configuration thanks to particle number fluctuations in the fragments. 
Nucleon transfer is expected to be stronger in the $yx$ configuration because, in addition to fluctuations, no spatial symmetry prevents from an average flux of nucleons from one nucleus to the other.
Indeed, integration of  proton and neutron densities in each reactant after the $yx$ collision indicates an average transfer of $\sim6$~protons and $\sim11$~neutrons from the right to the left nucleus.
In this case, transfer occurs from the tip of the aligned nucleus to the side of the other. 
The $yx$ configuration is then expected to favor the formation of nuclei heavier than~$^{238}$U. 

\begin{figure}
\includegraphics[width=8cm]{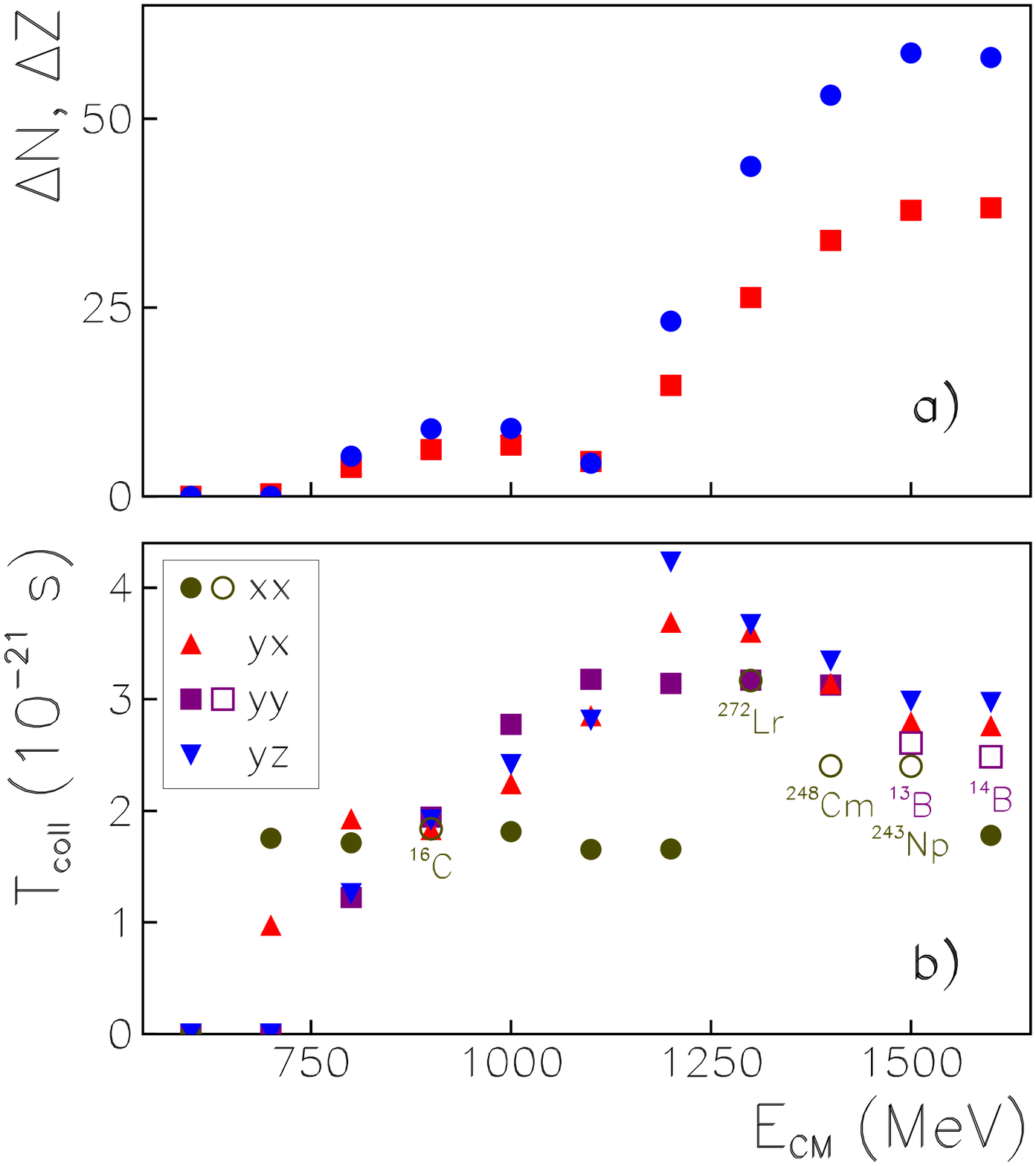}
\caption{a) Number of transfered protons (squares) and neutrons (circles) in the $yx$ configuration and b) collision times for each orientation as function of center of mass energy. Empty symbols indicate that three fragments are produced in the exit channel. In this case, the closest nucleus to the middle fragment is given. }
\label{fig:transfer}
\end{figure}

To get a deeper insight into this transfer, the role of collision energy is shown in Figure~\ref{fig:transfer}-a. 
Two regimes can clearly be identified. 
At $E_{CM} \leq 1000$~MeV, standard transfer dominates and nuclei up to $^{253}$Cf are produced from the average flux of nucleons (neglecting contributions from particle number fluctuations). 
In this low energy regime, the heavy fragment may survive to fission but is not expected to reach the SHE island of stability. 
Note that this is consistent with experiment as no transfermium nuclei  have been observed in this reaction and energy range~\cite{sch78}.
Here, single particle wave functions are transferred through the neck between the fragments with a smooth change of their shapes. 
In particular, the particle density in the neck increases with energy but is always lower than the saturation density $\rho_0=0.16$~fm$^{-3}$. 
This value is reached only at $E_{CM}\sim1000$~MeV. 
At higher energies, however, the contact area gets more dense and saturation density is overcome in the neck. 
This modifies dynamically the breaking point of the giant system and the left fragment gains much more nucleons than expected in standard transfer. 
Superheavy fragments up to $Z\sim130$ and $N\sim205$ could be produced at $E_{CM}\sim1500$~MeV. 
However, at these energies, the heavy fragment is expected to have a very high temperature in such a way that it decays spontaneously into fission. 
Note that, for such violent collisions where densities well above saturation density are reached, nucleon-nucleon collisions might play a role on top of the mean field evolution and extensions of TDHF including collision term should be considered to check these predictions~\cite{asi09,lac04,toy02}. 


\begin{figure}
\includegraphics[width=8cm]{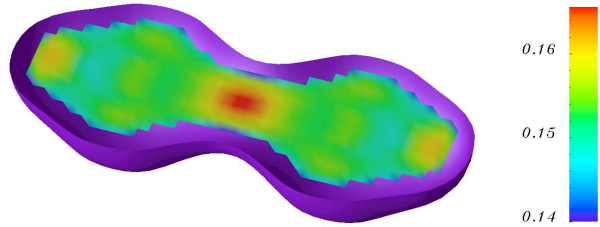}
\caption{Nucleon density (in fm$^{-3}$) in the collision plane is plotted when the density in the neck reaches its maximum in the $xx$ configuration at $E_{CM}=900$~MeV. The half cut surface is an isodensity at half the saturation density, i.e. $\rho_0/2=0.08$~fm$^{-3}$.}
\label{fig:rho_neck}
\end{figure}

Figure~\ref{fig:densities} also shows that, in the $xx$ configuration, the giant system breaks in three fragments. 
This was a surprise as a static macroscopic approach predicted this phenomenon to be strongly hindered as compared to binary fission~\cite{wu84}. Here, integration of proton and neutron densities indicates a $^{16}$C-like fragment. Why the system decides to form a third fragment instead of breaking at the neck?
The answer is given by a close look at its internal density. Figure~\ref{fig:rho_neck} shows a snapshot of the density in the collision plane obtained in the $xx$ configuration at~$E_{CM}=900$~MeV
at closest approach.
Fragment tips strongly overlap and the density reaches a maximum of~$0.166$~fm$^{-3}$ in the neck, exceeding saturation density~$\rho_0$.
This places the system in the ternary fission valley of the potential energy 
surface. To much charges are present in the center and the Coulomb energy 
is not efficient to compensate for the surface energy increase which induces the 
formation of two necks~\cite{wu84}.
The same phenomenon is observed in the $yy$ configuration at $E_{CM}=1500$ and 1600~MeV. 
Here, the middle fragment corresponds to a neutron rich boron isotope (see Figure~\ref{fig:transfer}-b). 
This occurs at higher energy than in the $xx$ case because a closer distance is needed to overcome the saturation density in the neck. 
(To reach closer distances, the nuclei need more energy as the Coulomb repulsion get stronger.) 
Note that overcoming saturation density is not a sufficient condition to produce three fragments as, e.g., such an exit channel is not observed in the $yz$ configuration. 
A third (heavy) fragment is also produced at high energy in the $xx$ configuration (see Figure~\ref{fig:transfer}-b). 
In this case, however, the dynamics is driven by strong density fluctuations similar to the high energy regime in the $yx$ configuration (see Figure~\ref{fig:transfer}-a and above discussion on transfer).  

Let us finally investigate a last important aspect of collision dynamics, which is the collision time between nuclei. Here, the collision time is defined as the time during which the neck density exceeds $\rho_0/10=0.016$~fm$^{-3}$. 
Figure~\ref{fig:transfer}-b shows the evolution of collision time $T_{coll}$ as a function of $E_{CM}$ for each configuration.
In the low energy part ($E_{CM}\leq900$~MeV),  three distinct behaviors between the $xx$, $yx$ and $yy/yz$ configurations are seen.  In particular, the last need more energy to get into contact as the 
energy threshold above which nuclear interaction plays a significant role is higher for such compact configurations. 

Looking now at the whole energy range, the $yx$, $yy$ and $yz$ orientations exhibit roughly the same behavior, i.e., a rise and fall of $T_{coll}$ with a maximum of $3-4\times10^{-21}$~s at $E_{CM}\sim1200$~MeV. This position of the maximum is in agreement with the QMD calculations of Ref.~\cite{tia08}. 
Dynamical evolution of nuclear shapes in these three configurations, in addition to a strong transfer in the~$yx$ one (see Figure~\ref{fig:transfer}-a) are responsible for these rather long collision times as compared to scattering with frozen shapes of the reactants~\cite{zag06}. 
The~$xx$ configuration, however, behaves differently.
In this case, $T_{coll}$ exhibits a plateau which does not exceed~$2\times10^{-21}$~s except when a third heavy fragment is formed due to dynamical density fluctuations. 
This overall reduction of $T_{coll}$ in the~$xx$ case is attributed to the strong overlap of the tips. 
Indeed, as it can be seen at $E_{CM}=900$~MeV in Figure~\ref{fig:rho_neck}, this overlap produces a density in the neck higher than saturation density. 
The fact that nuclear matter is difficult to compress translates into a strong repulsive force between the fragments which decreases their contact time. 
This phenomenon is also responsible for the fall of collision times in the other configurations, though occurring at higher energies due to the fact that closer distances between the reactants are needed to strongly overlap.

To conclude, this fully microscopic quantum investigation of collision dynamics of two uranium nuclei exhibits a rich phenomenology which is strongly influenced by the shape  of the atomic nuclei. 
Let us summarize the three main conclusions of this study. 
($i$) The giant system formed in bare uranium-uranium  central collisions is expected to 
survive enough time to allow experimental observation of spontaneous 
positron emission, but only for some initial orientations (which reduces slightly the cross section of the process) and with an energy $E_{CM}\geq1000$~MeV. 
($ii$) Heavy fragments produced at low energy might survive fission, but the center of their charge distribution is not expected to reach the transfermium region. This  indicates that this reaction might not be appropriate to populate the blank spot region of the nuclear chart between decay chains of superheavy elements produced by ''hot'' and ''cold'' fusion. 
These exploratory quantum calculations have been possible due to recent increase of computational power. We then expect more systematic studies of other actinide collisions in a near future to look for optimized channels for transfermium and SHE productions, making use, for instance, of the "inverse quasi-fission" process~\cite{zag06}.
($iii$) Experimental observation of carbon-like fragments at $E_{CM}=900$~MeV would sign ternary fission.
We encourage experimental and theoretical investigations of this mechanism as 
it is an excellent probe to test the physical ingredients of this dynamical nuclear 
many-body approach. In particular, how is it affected by the different terms of  the
energy density functional and what is the effect of pairing are open questions.

We thank P. Bonche for providing his code.
We are also grateful to B. Avez, D. Boilley, R. Dayras, W. Greiner, S. Heinz
and A. Villari for discussions and a careful reading of the paper. 
The calculations have been performed in the 
Centre de Calcul Recherche et Technologie of the Commissariat \`a l'\'Energie Atomique.

\end{document}